\documentclass[%
 reprint,
 amsmath,amssymb,
 aps,
 showkeys,
]{revtex4-1}

\usepackage{graphicx}
\usepackage{dcolumn}
\usepackage{bm}

\begin{document}

\title{An Optimum Algorithm for Quantum Search}
\author{Jiang Liu}
\email{jiangliu@alumni.cmu.edu}%
\affiliation{%
 Scarsdale, NY 10583, USA
}%

\date[]{September 2nd, 2019}

\begin{abstract}
This paper discusses an improvement to Grover's algorithm for searches where target states are Hamming weight eigenstates and search space is not ordered. It is shown that under these conditions search efficiency depends on the smaller number of 0's and 1's, not the total length, of binary string of target state, and that Grover's algorithm can be improved whenever number of 0's and number 1's are not equal. In particular, improvement can be exponential when number of 0's or number of 1's is very small relative to binary string length. One interesting application is that Dicke state preparation, which in Grover's algorithm is P on average, can be made poly-efficient in all cases. For decision making process, this improvement won't improve computation efficiency, but can make implementation much simpler.
\end{abstract}

\maketitle


\section{Introduction}

Quantum search is a process of looking for a target state $\vert k\rangle$, where $0 \le k < 2^n$ and $n$ is the number of qubits of the system, out of a mixture of $2^n$ un-ordered states 
\begin{eqnarray}
&&	\vert \psi_{\theta}\rangle
	\equiv 
	\frac{1}{\sqrt{2^{n}}}\sum_{j=0}^{2^n-1}\vert j\rangle \longrightarrow \vert k  \rangle	
\label{eqn:QuSearch}\\
&&\sin\theta =  \frac{1}{\sqrt{2^n}}	
\label{eqn:BasisMixing0}
\end{eqnarray}

Efficiency of quantum search plays an important role in the design of collision-resistance cryptography\cite{katz08,Kim2018,Hosoyamada2019,Brassard1997}.

In the absence of coherent interference all states in the mixture are independent. Number of trials of success of finding $\vert k\rangle$ on average is the same as that in classical search, for which query complexity is $\bm{O}(2^n)$. With coherent interference in quantum computing, Grover\cite{Grover} discovered that search efficiency can be sped-up quadratically, reducing query complexity to $\bm{O}(2^{n/2})$.

Grover's algorithm is generated, in part, by rotating qubit along polar axes of Bloch sphere\cite{Bloch46} by $\pi/2$.
Qubit rotation  plays an important role in quantum computing\cite{NC2010}. The most famous example so far is Quantum Fourier Transformation (QFT)\cite{Deu85,Cop94,GN96,CEMM98,NC2010} and its applications such as Shor's algorithm{\cite{Shor94} for factorization and   Harrow-Hassidim-Lloyd algorithm\cite{HHL2009} for certain linear system of equations. QFT is generated by rotating qubit along azimuthal axes of Bloch sphere. 
	
It is well known that Grover's algorithm is optimum \cite{Zalka99,Bennett97} when qubit polar angle is $\pi/2$, i.e., when Hadamard matrix is unbiased. It is, however, less clear what would happen if Hadamard matrix is biased.  The idea of using biased Hadamard transformation is not new\cite{Childs2000}. The purpose of this paper is to explore if Grover's algorithm can be improved under certain conditions when Hadamard transformation is biased.

In section II,  dynamics of Grover's algorithm is recast to oscillation. Although Grover's algorithm can be viewed from different angles \cite{Bennett97,Boyer98,Brassard2002}, viewing from oscillation angle \cite{Pont58,MNS1962,Wolfenstein78,MSW85} can provide useful insight into where and how improvement may be made.  

Following oscillation view, section III discusses an optimum search algorithm, referred to as Grover-Plus henceforth for convenience, for Grover's original database search, where both target $\vert k\rangle$ and its oracle are given but search list $\vert \psi_{\theta}\rangle$ is unordered. In this case, the only difference between Grover-Plus and Grover's algorithm is that qubit polar angle is a variable of Hamming weight\cite{Hamming1950} and that polar angle is determined by optimization.

It will be shown that while Grover-Plus and Grover's algorithm are identical when Hamming weight of target state equals $n/2$, Grover-Plus is always more efficient in any other cases. In particular, Grover-Plus is exponentially more efficient when Hamming weight of target state is either very small or very close to $n$. 

Section IV discusses an interesting application of Grover-Plus: Dicke states\cite{Dicke1954} production.  Dicke states play a special role in quantum entanglement and quantum computing (for details see \cite{Childs2000,Ionicioiu2008,Bergmann2013,Chakraborty2013, Bartschi2019} and references therein). Grover-Plus turns out to be simpler and  more efficient than existing methods
\cite{Childs2000,Ionicioiu2008,Chakraborty2013, Bartschi2019}. Query complexity of Grover-Plus is $\bm{O}\left((\Delta_{0,k}(1-\Delta_{0,k}/n))^{1/4}\right)$, where $\Delta_{0,k}$ is the Hamming weight of Dicke state.

Section V compares Grover-Plus to Grover searches where searching space can be limited to states of equal Hamming weight. In this case, Grover's algorithm and Grover-Plus have similar computation efficiency. Implementation of Grover-Plus turns out to be much simpler.

Conclusion of this paper is summarized in section VI. Some details of calculation are organized in the Appendix.

\section{Quantum Search and Quantum Oscillation}

Casting quantum search to quantum oscillation has two advantages: First,  it makes the proof of completeness of Grover's search \cite{Bennett97,Boyer98,Brassard2002} much simpler. Second, it openly reveals which part of Grover's algorithm is already optimized and where else it can potentially be improved. 

Let's start by re-deriving Grover's algorithm from oscillation point of view.

The $2^n$ dimensional Hilbert space of Eq.~(\ref{eqn:QuSearch}) can be partitioned to a coarse-grained two dimensional subspace supported by search target state $\vert k\rangle$ and its negation $\vert \tilde{k}\rangle$, defined by  
$\vert \tilde{k}\rangle \langle \tilde{k}\vert = 1- \vert k\rangle\langle k\vert$,
\begin{equation}
\vert\psi_{\theta}\rangle 
= \cos\theta\vert \tilde{k}  \rangle + \sin\theta\vert k\rangle,
\label{eqn:QuSearch2}
\end{equation}
where $\vert \psi_{\theta}\rangle$ and  $\theta$  are given by Eqs.~(\ref{eqn:QuSearch}) and (\ref{eqn:BasisMixing0}).

Oscillation takes place between  $(\vert \tilde{k}\rangle, \vert k\rangle)^T$ and  
$(\vert 0\rangle, \vert \tilde{0}\rangle)^T$ bases connected by
 \begin{equation}
 H^{\otimes{n}}
 \begin{pmatrix}
 \vert \tilde{k}\rangle\\
 \vert k\rangle
 \end{pmatrix}
 = U^{\dagger}
 \begin{pmatrix}
 \vert 0\rangle\\
 \vert \tilde{0}\rangle
 \end{pmatrix},
 \label{eqn:AntiMixing}
 \end{equation}
 where $\vert \tilde{0} \rangle$ is the negation of 
 $\vert 0\rangle$, $H$ 
 is Hadamard gate and 
 \begin{equation}
 U_{\theta} = 
 \begin{pmatrix}
 \cos\theta  & \sin\theta\\
 -\sin\theta  & \cos\theta
 \end{pmatrix}
 \label{eqn:GMixingMatrix}
 \end{equation}
 is a bases mixing matrix. That $(\vert 0\rangle, \vert \tilde{0}\rangle)^T$ comes into play is because $\vert \psi_{\theta}\rangle = H^{\otimes{n}}\vert 0\rangle$.
 
Evolution of $(\vert \tilde{k}\rangle, \vert k\rangle)^T$  is given by
\begin{equation}
\begin{pmatrix}
\vert \tilde{k}\rangle \\
\vert k\rangle
\end{pmatrix}_{t}
=
\left( T_{\theta} \right)^t
\begin{pmatrix}
\vert \tilde{k}\rangle \\
\vert k\rangle
\end{pmatrix}_0
\label{eqn:GEvolution},
\end{equation}
where $t$ is the number of oscillation cycles and 
\begin{equation}
T_{\theta} 
=
U_{\theta}
\begin{pmatrix}
1 & 0\\
0 & -1
\end{pmatrix}
U_{\theta}^{\dagger}
\begin{pmatrix}
-1 &  0 \\
0        & 1
\end{pmatrix}.
\label{eqn:GEvolutionGates}
\end{equation}

Diagonal matrices of $T_{\theta}$ are referred to as Grover's oracle in the literature. They are oscillation matrices of bases eigenvectors. That they are traceless implies that oscillation is maximum. One important consequence of maximum oscillation is
\begin{equation}
\left( T_{\theta} \right)^t = T_{t\theta}.
\label{eqn:Coherent}
\end{equation}
resulting in
\begin{equation}
Pr\left( k\vert \psi_{\theta} \right)_{t} = \sin^2\left(\theta + 2t\theta\right).
\label{eqn:GProbability}
\end{equation}

This is Grover's result re-derived from the oscillation point of view. Query complexity of Eq.~(\ref{eqn:GProbability}) is determined by  $t = \lfloor(\pi/\theta-2)/4\rfloor = \bm{O}(2^{n/2})$.

Dynamics of Grover's algorithm is similar to that of neutrino oscillation\cite{Wolfenstein78,MSW85,Smirnov13} in that both involve bases mixing and oscillation. In particular,  $(\vert 0\rangle,\vert \tilde{0}\rangle)^T$ and $(\vert \tilde{k}\rangle, \vert k\rangle)^T$ 
play similar roles as mass and interaction bases\cite{Wolfenstein78} in neutrino oscillation.  In Grover's algorithm, mixing and oscillation take place at different time and location. By contrast, mixing and oscillation of neutrinos in interaction-bases are modeled at the same time and at the same location.

Oscillation efficiency is driven by three factors\cite{Pont58,Wolfenstein78,MSW85,Smirnov13}: (1) structure of oscillation matrix, (2) bases-mixing and (3) choice of oscillation bases. In view of oscillation physics, Grover's algorithm is already optimum with respect to the structure of oscillation matrix. However, issues related to bases-mixing and oscillation  bases were not addressed. These are the areas where improvement may potentially be made.


\section{A Generalization of Grover's Algorithm}
\label{Oscillation in Binary System}

Following oscillation view, Grover's computation model \cite{Grover,Brassard97,Boyer98,Brassard2002}  may be generalized by allowing arbitrary polar angle and arbitrary intermediate bases, while keeping maximum oscillation feature unchanged, i.e.,
\begin{eqnarray}
&&\begin{pmatrix}
\vert 0\rangle \\
\vert \tilde{0}\rangle
\end{pmatrix}
 \rightarrow
\begin{pmatrix}
\vert j\rangle \\
\vert \tilde{j}\rangle 
\end{pmatrix},
	\label{eqn:IntermediateBasis}\\
&&H \rightarrow H_{\zeta} = \sigma_z\cos\frac{\zeta}{2} + \sigma_x\sin\frac{\zeta}{2},
\label{eqn:QubitRotation}
\end{eqnarray}
where $\sigma_x$ and $\sigma_z$ are the Pauli matrices, $H_{\zeta}$ is a generalized Hadamard gate with $0\le \zeta\le \pi$.
Changing intermediate basis is not necessary for the optimum search algorithm, Grover-Plus, discussed below, but is indispensable for algorithms requiring a multiple-step evolution path. 

Optimizing bases-mixing involves two parts. First, determining a polar angle $\zeta$ that optimizes bases-mixing. Second, setting an initial condition that connects optimized oscillation bases to $\vert \psi_{\theta}\rangle$ so that results of optimization can be applied to searches described by Eq.~(\ref{eqn:QuSearch}). 

Let's start from the first part. Operationally, $H_{\zeta}^{\otimes{n}}$  turns a pure state $\vert j\rangle$ to a superposition. Weight of $\vert k\rangle$ in the superposition represents mixing between $\vert j\rangle$ and $\vert k\rangle$. It is shown in the Appendix that  $\vert j\rangle$ and $\vert k\rangle$ mixing depends on $\zeta$ and on number of bit-flippers between $\vert j\rangle$ and $\vert k\rangle$. If binary strings of $\vert j\rangle = \vert j_n\cdots j_2j_1\rangle$ and $\vert k \rangle = \vert k_n\cdots k_2k_1\rangle$ are viewed as members of ordered set, number of bit-flippers is their Hamming distance\cite{Hamming1950}
\begin{equation}
\Delta_{j,k}  = \sum_{\alpha=1}^{n}\left( j_{\alpha} - k_{\alpha} \right)^2. \label{eqn:SymmetricDifference}
\end{equation}
When $j=0$, $\Delta_{j,k}=\Delta_{0,k}$ becomes the Hamming weight of $\vert k_n\cdots k_2k_1\rangle$.

If target state is an eigenstate of $\Delta_{j,k}$, optimum solution to polar angle $\zeta$ turns out to be (see Appendix for detail)
\begin{equation}
\sin^2\frac{\zeta}{2}	= \frac{\Delta_{j,k}}{n},
\label{eqn:OptimalQubitMixing}
\end{equation}
and 
\begin{equation}
\sin^2\theta_{\Delta_{j,k}} = 
\left(\frac{\Delta_{j,k}}{n}\right)^{\Delta_{j,k}}
\left(1-\frac{\Delta_{j,k}}{n}\right)^{n-\Delta_{j,k}},
\label{eqn:OptimalBasisMixing}
\end{equation}
where $\theta_{\Delta_{j,k}}$ is the mixing angle between $(\vert j\rangle,\vert \tilde{j}\rangle)^T$ and $(\vert \tilde{k}\rangle, \vert k\rangle)^T$ bases, and that probability of finding $\vert k\rangle$ from $\vert \psi_{\theta_{\Delta_{j,k}}}\rangle$ after $t$ cycles of oscillation is
\begin{equation}
Pr\left( k\vert \psi_{\theta_{\Delta_{j,k}}} \right)_{t} 
= 
\sin^2\left(\theta_{\Delta_{j,k}} + 2t\theta_{\Delta_{j,k}}\right),
\label{eqn:GProbability2}	
\end{equation}
where $\vert \psi_{\theta_{\Delta_{j,k}}}\rangle = H^{\otimes{n}}_{\zeta}\vert 0\rangle$ is the analog of 
$\vert \psi_{\theta}\rangle$ of Eq.~(\ref{eqn:QuSearch2}). 

Notice, polar angle described by Eq.~(\ref{eqn:OptimalQubitMixing}) is exactly the same as that suggested, heuristically, by \cite{Childs2000} for Dicke state \cite{Dicke1954} production. This paper shows that the choice of \cite{Childs2000} is optimal.

To apply these results to Grover's search space, one must connect $\vert \psi_{\theta_{\Delta_{j,k}}}\rangle$ to 
$\vert \psi_{\theta}\rangle$. This is handled by the second part of Grover-Plus discussed below.

Permissible initial states in generalized Grover's computation model are
$H^{\otimes{n}}_{\zeta'}\vert 0\rangle$ for arbitrary $\zeta'$. 
Among them, $H^{\otimes{n}}_{\zeta}\vert 0\rangle$  with $\zeta$ determined by  Eq.~(\ref{eqn:OptimalQubitMixing}) has largest mixing with $\vert k\rangle$.
$\zeta$ is also the same angle that maximizes oscillation.
The natural choice of initial state for 
$\vert k\rangle$
is therefore 
$H^{\otimes{n}}_{\zeta}\vert 0\rangle$. All other initial states can be transformed to $H^{\otimes{n}}_{\zeta}\vert 0\rangle$
by an unitary base-shift operator $H^{\otimes{n}}_{\zeta}H^{\otimes{n}}_{\zeta'}$.

To apply Eqs.~(\ref{eqn:OptimalQubitMixing}) and (\ref{eqn:OptimalBasisMixing}) to Grover's search space,  one simply needs to add a base-shift operator ahead of oscillation, i.e., 
\begin{equation}
\left(T_{\theta}\right)^t \rightarrow \left(T_{\theta_{\Delta_{0,k}}}\right)^{t}H^{\otimes{n}}_{\zeta}H^{\otimes{n}} = 
T_{t\theta_{\Delta_{0,k}}}H^{\otimes{n}}_{\zeta}H^{\otimes{n}}.
\label{eqn:coherent3}
\end{equation}
Probability of finding $\vert k\rangle$ from $\vert \psi_{\theta}\rangle$ after $t$ cycles of oscillation now becomes
\begin{equation}
Pr\left( k\vert \psi_{\theta} \right)_{t} 
= 
\sin^2\left(\theta_{\Delta_{0,k}} + 2t\theta_{\Delta_{0,k}}\right).
\label{eqn:GProbability3}	
\end{equation}

Eqs.~(\ref{eqn:OptimalQubitMixing}), (\ref{eqn:OptimalBasisMixing}), (\ref{eqn:coherent3}) and (\ref{eqn:GProbability3}) define Grover-Plus when target state is an eigenstate of Hamming weight and search space is given by Eq.~(\ref{eqn:QuSearch}). They are the major results of this paper. They have the following interesting features.

When $\Delta_{0,k}=n/2$, Grover-Plus and Grover's algorithm are identical.  In this case, $\zeta=\pi/2$ and  $\sin\theta_{n/2} = \sin\theta = 2^{-n/2}$. In other words, Grover's algorithm is a special case of  Grover-Plus,  and is optimum when $\Delta_{0,k}=n/2$. 

When $\Delta_{0,k} \ne n/2$, Grover-Plus is more efficient than Grover's algorithm because bases-mixing in Grover-Plus is larger than that in Grover's algorithm.
Asymptotically, one can show from Eq.~(\ref{eqn:OptimalBasisMixing}) that
\begin{equation} 
\theta_{\Delta{_{j,k}} }\sim
\begin{cases}
\left(\delta_{j,k}/n\right)^{\delta_{j,k}/2}   
& \quad \text{when } \delta_{j,k}\ll n/2 \\
\left(\delta_{j,k}/n\right)^{\delta_{j,k}}
& \quad \text{when } \delta_{j,k} \sim n/2
\end{cases}
\label{eqn:hierarchy}
\end{equation}
where 
\begin{equation}
\delta_{j,k}\equiv \min\left[\Delta_{j,k},n-\Delta_{j,k}\right],
\label{eqn:MinBitFlipper}
\end{equation}
is the smaller of $\Delta_{j,k}$ and $n-\Delta_{j,k}$. $\delta_{j,k} = n/2$ corresponds to the case for Grover's algorithm.

Query complexity of Grover-Plus is determined by 
 $t = \lfloor(\pi/\theta_{\Delta_{0,k}})-2)/4\rfloor$, which leads to
 $\bm{O}\left(c^{n/c}\right)$, when $\delta_{j,k} \sim n/c$ where $c \ge 2$,  and $\bm{O}\left(n^{\delta_{j,k}/2}\right)$ when $\delta_{j,k} \ll n/2$. In all these cases, Grover-Plus is always more efficient than  Grover's algorithm whose query complexity is always equal to $\bm{O}\left(2^{n/2}\right)$.

In particular, when $\delta_{j,k} \ll n$, query complexity of Grover-Plus is polynomial, which is exponentially more efficient than Grover's algorithm. 

Of special interest are cases where target is  $\vert 0\rangle$ or 
$\vert 2^n-1\rangle$. 
In these cases $\Delta_{0,k}=0$ or $n$. Grover-Plus solution is $\zeta = 0$ or $\pi$, $\theta_{\Delta_{0,0}}= \theta_{\Delta_{0,2^n-1}} = \pi/2$ and  $t = 0$.  No oscillation is needed, i.e., $t=0$, base-shift  $H^{\otimes{n}}_{\zeta}H^{\otimes{n}}$ alone can do the job. In Grover's algorithm, by contrast, one has to query Grover's Oracle $2^{n/2}$ times to get the same results.

Another interesting example is the process
where $\delta_{0,k}=1$. Grover-Plus solution is $\sin^2\frac{\zeta}{2}=n^{-{1/2}}$ or $1-n^{-1/2}$, query computational complexity is $\bm{O}\left(\sqrt{n}\right)$. Compared to  Grover's algorithm, speed-up of Grover-Plus is exponential. 

Implementation of Grover-Plus is straightforward.  Once search target $\vert k\rangle$ is specified, Hamming weight of $\vert k\rangle$ is known. All one needs to do is (1) dial qubit polar angle to that described by Eq.~(\ref{eqn:OptimalQubitMixing}), (2) add base-shift gate in front of oscillation oracles, and (3) make measurement after $\lfloor(\pi/\theta_{\Delta_{0,k}}-2)/4\rfloor$ queries.
 
\section{Dicke State Preparation}

This section discusses an interesting application of Grover-Plus: Dicke state preparation. A Dicke state is an equally weighted superposition of all states of the same Hamming weight
\begin{equation}
\vert D^n_{\Delta_{0,k}}\rangle \equiv {{n\choose \Delta_{0,k}}^{-1/2}}
\sum_{j}\vert j\rangle. \label{eqn:SpinSearch}
\end{equation}
Producing a Dicke state from $\vert 00\cdots 0\rangle$  is equivalent to Hamming weight search in Grover's unsorted search space $\vert \psi_{\theta}\rangle$. 

Classically,
probability density of finding a $\Delta_{0,k}$ in an unstructured dataset  is
\begin{equation}
\rho\left(\Delta_{0,k}\right) = 2^{-n}  {n \choose\Delta_{0,k}}.
\label{eqn:AverageClassical}
\end{equation}
Computational complexity of Eq.~(\ref{eqn:AverageClassical}) varies from  worst case of $\bm{O}(2^n/n)$, which is NP informally,  when $\delta_{0,k} \ll n/2$, where $\delta_{0,k}$ is given by Eq.~(\ref{eqn:MinBitFlipper}),  to  $\bm{O}(\sqrt{n})$, which is P, when $\delta_{0,k} \to n/2$. On average\cite{Levin1986,Sanjeev09}, this process is P because states with $\delta_{0,k}=\Delta_{0,k} = n/2$ are most populous.

Grover's algorithm can provide a quadratical speedup to classical search,but cannot change the nature of computational complexity.
 
Grover-Plus is applicable to $\vert D^n_{\Delta_{0,k}}\rangle$ search.  The only changes are (1) Grover's Oracle acts on all member state of $\vert D^n_{\Delta_{0,k}}\rangle$ equall, and (2) replacing bases-mixing of a pure state, $\theta_{\Delta_{0,k}}$, by bases-mixing of
 $\vert D^n_{\Delta_{0,k}}\rangle$, $\theta^n_{\Delta_{0,k}}$. One can show
 \begin{equation}
\sin^2\theta^n_{\Delta_0,k} = {n\choose \Delta_{0,k}}\sin^2\theta_{\Delta_{0,k}}.
 \label{eqn:Scaling}
 \end{equation}
 
Hamming-weight search is in  NP  only when $\delta_{0,k} \ll n/2$. This is exactly the same place where Grover-Plus is exponentially more efficient than Grover's algorithm. As a result, we expect that Grover-Plus is able to make Hamming-weight search poly-efficient for all cases.

Indeed, from  Stirling approximation
\begin{equation}
{n \choose\Delta_{0,k}}
\approx
\left[2\pi\Delta_{0,k}\left(1-\frac{\Delta_{0,k}}{n}\right)\right]^{-1/2}
\sin^{-2}\theta_{\Delta_{0,k}}\
\label{eqn:Stirling}
\end{equation}
where 
$\sin^2\theta_{\Delta_{0,k}}$ 
is given by Eq.~(\ref{eqn:OptimalBasisMixing}), one finds 
\begin{equation}
\sin^2\theta^n_{\Delta_{0,k}}
 \approx \left[2\pi\Delta_{0,k}\left(1-\frac{\Delta_{0,k}}{n}\right)\right]^{-1/2},
 \label{eqn:DickeStateMixing}
\end{equation}
and
\begin{equation}
Pr\left( D^n_{\Delta_{0,k}}\vert \psi_{\theta} \right)_{t} 
\approx
\sin^2\left(\theta^n_{\Delta_{0,k}} + 2t\theta^n_{\Delta_{0,k}}\right),
\label{eqn:GP4HammingWeight}	
\end{equation}
where approximation follows from  Eq.~(\ref{eqn:Stirling}). Computational complexity of Eqs.~(\ref{eqn:DickeStateMixing}) and (\ref{eqn:GP4HammingWeight}) is determined by  $t=\lfloor (\pi/\theta_{\Delta^n_{0,k}}-2)/4\rfloor = \bm{O}\left((\Delta_{0,k}(1-\Delta_{0,k}/n))^{1/4}\right)$, which is 
in P for arbitrary $\Delta_{0,k}$.
 
As far as Dicke state production is concerned, Grover-Plus is a simple realization of computation model hypothesized in \cite{Childs2000}. It provides a simple close-form description of Dicke state wave function of arbitrary Hamming weight at all time. In particular, it shows that the choice of biased Hadamard angle in \cite{Childs2000} is optimal (see Eqs.~(\ref{eqn:OptimalQubitMixing})) and (\ref{eqn:OptimalBasisMixing})).

Grover-Plus is poly-efficient in $\min[n,n-\Delta_{0,k}]$, while others 
\cite{Childs2000,Ionicioiu2008,Chakraborty2013,Bartschi2019} are poly-efficient in $n$. Query complexity of these models, whenever available, vary from  $\bm{O}(n),~\bm{O}(n^{1/2})$ to $\bm{O}(n^{1/4})$. Grover-Plus is likely to be more efficient when $\Delta_{0,k} \ne 0,n, n/2$, because $\bm{O}\left((\Delta_{0,k}(1-\Delta_{0,k}/n))^{1/4}\right) \le \bm{O}(n^{1/4})$. Grover-Plus does not require ancilla qubits.

\section{Searching States of Equal  Hamming Weight}

When Hamming weight of the searching target is known, an alternative strategy is to use Grover's algorithm and limit searching space to states whose Hamming weight are equal to that of the target state. For convenience, we refer to this strategy as modified Grover's algorithm. Modified Grover's algorithm is often used in decision making process as an equation solver. In this section, we are to compare Grover-Plus to modified Grover's algorithm.

Modified Grover's algorithm requires two conditions absent in Grover-Plus. One  is that space $\left\{\vert k\rangle\right\}, k = 0,1,\cdots, 2^n-1$, is ordered. The other is that qubit rotation and Grover's Oracle act on different spaces.

The first condition is to makes sure that selection of equal Hamming weight states, $\left\{\vert \Delta_{0,k}\rangle\right\}$, from $\left\{\vert k\rangle\right\}$ is possible and efficient. Here $\left\{\vert \Delta_{0,,k}\rangle\right\}$ represents states whose Hamming weight is $\Delta_{0,k}$.  When $\Delta_{0,k} =1$, for example,  member states of $\left\{\vert \Delta_{0,K}\rangle\right\}$ are $\vert 0\cdots  01\rangle, \vert 0\cdots 10\rangle, \cdots,\vert 1\cdots 00\rangle$ in binary representation and $\vert 2^k\rangle$, $k=0,1,\cdots,n-1$, in digital representation.

The second condition rises because $\left\{\vert\Delta_{0,k}\rangle\right\}$ is not a Hilbert subspace of $\left\{\vert k\rangle\right\}$ with algebra $H^{\otimes n}_{\zeta}$. $H_{\zeta}^{\otimes n}\vert \Delta_{0,k}\rangle$ will bring in states outside of $\left\{\vert \Delta_{0,k}\rangle \right\}$. As a consequence,  applying Eq.(\ref{eqn:GEvolutionGates}) to  $\left\{\vert \Delta_{0,k}\rangle \right\}$ will not produce required coherent interference for amplitude amplification.

To work round, one can introduce a working registry of $n'$ qubits, $\left\{\vert k'\rangle\right\}, k' = 0,1,\cdots, 2^{n'}-1$, and map $\left\{\vert \Delta_{0,k}\rangle\right\}$ to $\left\{\vert k'\rangle\right\}$ (with potentially trivial padding states in $\vert k'\rangle$ )
\begin{equation}
\left\{\vert \Delta_{0,k}\rangle\right\} \longleftrightarrow  \left\{\vert k'\rangle\right\}.
\label{eqn:RegistryMapping}
\end{equation}
Size of $\left\{\vert \Delta_{0,k}\rangle \right\}$ is binormially distributed. Required number of qubits of working registry, $n'$, is determined by
\begin{equation}
{n \choose \Delta_{0,k} +1} > 2^{n'} \ge {n \choose \Delta_{0,k}}.
\label{eqn:working_registry}
\end{equation}
In this implementation, Grover's Oracle is acting on $\left\{\vert k\rangle\right\}$ as before, Grover's phase shift, $\vert 0\rangle \to -\vert 0\rangle$,  qubit polar rotation $H^{\otimes n'}$ and final measurement are  on working registry $\left\{ \vert k'\rangle\right\}$.

In modified Grover's algorithm, bases mixing is determined by
\begin{equation}
\sin\theta' = \frac{1}{\sqrt{2^{n'}}},
\label{eqn:BasesMixing3}
\end{equation}
which can be orders of magnitude bigger than that in Grover algorithm if $n'\ll n$. Accordingly, modified Grover's algorithm can be orders of magnitude more efficient than Grover's algorithm.

Comparing modified Grover algorithm to Grover-Plus, we have from Eqs.(\ref{eqn:Stirling}), (\ref{eqn:working_registry}) and (\ref{eqn:BasesMixing3})
\begin{equation}
\frac {\sin\theta'}{\sin\theta_{\Delta_{0,k}}} \sim
	\left[ 2\pi\Delta_{0,k}\left( 1-\frac{\Delta_{0,k}}{n}\right)\right]^{1/4},
\end{equation}
i.e., bases mixings of modified Grover's algorithm and Grover-Plus are of same order of magnitude for all Hamming weight. As a result,  modified Grover algorithm and Grover-Plus are equally efficient. 

Advantage of Grover-Plus is that its implementation is simpler.  Grover-Plus works in the Grover's database search Hilbert space $\left\{ \vert k \rangle\right\}$. It does not require $\left\{ \vert k \rangle\right\}$ be ordered, nor need an extra working registry to map to for every iteration.


\section{Conclusion}

Based on the observation that dynamics of quantum search is quantum oscillation, it is shown that Grover's database search algorithm can be improved when search target is a Hamming weight eigenstate and Hamming weight of the target state is known. 

Improvement rises from an optimum choice of qubit polar rotation angle determined by Hamming weight.
Magnitude of improvement varies from exponential when Hamming weight of target state is close to $0$ or $n$ to no improvement when Hamming weight is equal to $n/2$.  

As far as computational complexity is concerned, choosing optimum qubit rotation angle is in general equivalent to limiting searching space to states of equal Hamming weight. It is shown, however, that choosing optimum rotation angle is much easier to implement.

After the completion of this work, I noticed a very interesting work of A. Gilliam, M. Pistoia and C. Gonciulea \cite{Gonciulea20} on the same topic. Although our approaches and focuses are somewhat different, we independently reached the same results shown by 
Eqs.(\ref{eqn:OptimalQubitMixing}) and (\ref{eqn:OptimalBasisMixing}).

\begin{acknowledgments}
	I would like to thank Wei-Min Shen,  Greg Ver Steeg,  Jiang Yuan and Stuard Hadfield for valuable comments and suggestions, and to Edward M. Liu for  interesting discussions and encouragement.  All remaining errors are mine.

\end{acknowledgments}

This research did not receive any specific grant from funding agencies in the public, commercial, or not-for-profit sectors.

Declarations of interest: None.

Disclaimer:  views expressed in this paper are mine as an individual and not as a representative speaking for or on behalf of my employer, nor do they represent my employer's positions, strategies or opinions.

\appendix*

\section{Oscillation in Binary System}

Relation  between qubit polar rotation angle and bases mixing is most evident in binary presentation of quantum state, in which $\vert j\rangle$ can be written as
\begin{eqnarray} 
&&\vert j\rangle = \vert j_n\cdots j_2j_1\rangle, \label{eqn:BinaryRep}\\
&& j = \sum_{\alpha=1}^nj_{\alpha}2^{\alpha-1},   \label{eqn:BinaryNumber}
\end{eqnarray}
where $j_{\alpha} \in \{0,1\}$ and $\alpha = 1,2,\cdots n$. $j_n
\cdots j_2j_1$ is referred to as binary string of $j$ in literature. The length of $j_n
\cdots j_2j_1$ is $n$. In general, computational complexity of a process involving $\vert j\rangle$ is measured by the length of binary string. Depending on the detail of processes, the result could be polynomial, exponential or others.

In binary presentation, differences between $j_n\cdots j_2j_1$ and $ k_n\cdots k_2k_1$ may be measured  by  their intersection
\begin{equation}
I_{j,k} = \sum_{\alpha=1}^{n}j_{\alpha}k_{\alpha},
\label{eqn:Intersection}
\end{equation}
and symmetric difference $\Delta_{j,k}$, defined by Eq.~(\ref{eqn:SymmetricDifference}),  which counts the number of bit-flippers between 
$\vert j_n\cdots j_2j_1\rangle$ and
$\vert k_n\cdots k_2k_1\rangle$.  For example, when $j=0$, $\Delta_{0,k} = \sum_{\alpha}k_{\alpha}$ is the number of 1's in binary string of $\vert k\rangle$

From Eqs.~(\ref{eqn:QubitRotation}) to (\ref{eqn:SymmetricDifference}) and binomial expansion, it can be shown that
\begin{equation}
H_{\zeta }^{\otimes n} \vert j\rangle =  \sum_{k=0}^{2^n-1}
\left(-1\right)^{I_{j,k}} 
\sin\theta_{\Delta_{j,k}}
\vert k\rangle,
\label{eqno:GeneralHadamard}
\end{equation}
where
\begin{equation}
\sin\theta_{\Delta_{j,k}} = 
\left(\cos\frac{\zeta}{2}\right)^{n-\Delta_{j,k}}
\left(\sin\frac{\zeta}{2}\right)^{\Delta_{j,k}}.
\label{eqn:GeneralMixing}
\end{equation}

Oscillation bases involving $\vert j\rangle$ and $\vert k\rangle$ are  $(\vert j\rangle, \vert \tilde{j}\rangle)^T$ and 
$(\vert \tilde{k}\rangle, \vert k\rangle)^T$, where $\vert \tilde{j}\rangle$ and
$\vert \tilde{k}\rangle$ are the negations of $\vert j\rangle$ and $\vert k\rangle$, in the sense that  $\vert \tilde{k}\rangle \langle \tilde{k}\vert = 1- \vert k\rangle\langle k\vert$
and that $\vert \tilde{j}\rangle \langle \tilde{j}\vert = 1- \vert j\rangle\langle j\vert$, respectively. They are connected by
\begin{equation}
\begin{pmatrix}
\vert j\rangle\\
\vert \tilde{j}\rangle
\end{pmatrix}
=U_{\theta_{\Delta_{j,k}}} H_{\zeta}^{\otimes{n}}
\begin{pmatrix}
\vert \tilde{k}\rangle\\
\vert k\rangle
\end{pmatrix},
\label{eqn:AntiMixing2}
\end{equation}
where $U_{\theta_{\Delta_{j,k}}}$ is a generalization of Eq.~(\ref{eqn:GMixingMatrix}) with $\theta$ replaced by $\theta_{\Delta_{j,k}}$ of Eq.~(\ref{eqn:GeneralMixing}). Ignoring an over all phase of no physical significance, one can show
\begin{eqnarray}
&&\vert \tilde{j}\rangle =
\sum_{i\ne j}\left(-1\right)^{I_{i,k}-I_{j,k}}
\frac{\sin \theta_{\Delta_{i,k}}}{\cos\theta_{\Delta_{j,k}}}\vert i\rangle
\label{eqn:Tilde_J} \\
&&\vert \tilde{k}\rangle = 
\sum_{i\ne k}\left(-1\right)^{I_{j,i}-I_{j,k}}
\frac{\sin \theta_{\Delta_{j,i}}}{\cos\theta_{\Delta_{j,k}}}\vert i\rangle.
\label{eqn:BinaryNumber2}
\end{eqnarray}

For a given $\zeta$, $\vert j\rangle$ and $\vert k\rangle$,  
the $2^n$ dimensional Hilbert space can be partitioned into a coarse-grained
two-dimensional subspace, in which $H^{\otimes{n}}_{\zeta}\vert j\rangle$ is the
initial superposition $\vert \psi_{\theta_{\Delta_{j,k}}}\rangle$, $\vert k\rangle$ is the target to be searched.  Evolution of $(\vert \tilde{k}\rangle, \vert k\rangle)^T$ is similar to  that in Grover's algorithm with mixing  $\theta_{\Delta_{j,k}}$ determined by $\Delta_{j,k}$ and $\zeta$. Optimizing $Pr(k\vert \psi_{\theta_{\Delta_{j,k}}})$ with respect to $\zeta$ produces Eqs.~(\ref{eqn:OptimalQubitMixing}) and (\ref{eqn:OptimalBasisMixing}).

\nocite{*}

\bibliography{QuSearch11}

\end{document}